\documentclass{kapproc}
\usepackage{psfig}
\begin{document}
\articletitle[]{Chemical Evolution \\ of CNO Abundances}
\chaptitlerunninghead{Chemical Evolution of CNO Abundances}

\author{M. GAVIL\'{A}N AND M. MOLL\'{A}}
\affil{Departamento de F\'{\i}sica Te{\'o}rica \\
 Universidad Aut{\'o}noma de Madrid, Cantoblanco, E-28049 Madrid, Spain}
\begin{abstract}
New low and intermediate mass star yields 
recalculated by Buell (1987) are evaluated by using
them in a Galactic Chemical Evolution model.
We analyze their effects on CNO elemental abundances
\end{abstract}

\section{Introduction}

Stellar yields for low and intermediate mass (LIM) stars have been
recalculated by Buell (1997), who, taking into account the effects of the
convective dredge up and the Hot Bottom Burning (HBB), incorporates
the most recent improvements in the TP-AGB processes.

The aim of this job is to evaluate these new yields for LIM stars
by using them as input in a Galactic Chemical Evolution (GCE)
model. In particular, we analyze if the contribution of these
stars is enough to justify the observed behavior of nitrogen vs.
oxygen.

\section{Stellar structure for LIM stars during the TP-AGB phase}

\begin{figure}[t]
\psfig{file=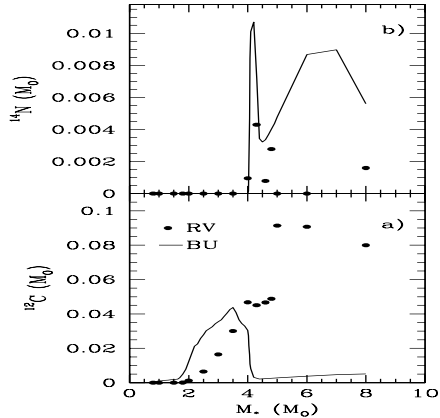,width=12cm,height=6cm}
\caption{Carbon and nitrogen ejected for LIM stars for $Z=0.02$. Solid
lines are BU yields, points are from RV.}
\label{yields}
\end{figure}

The evolution of a LIM star during the TP-AGB phase is dominated
mainly by two processes: The third dredge up events and the Hot Bottom
Burning (HBB). The first ones affect stars with masses between 2
M$_{\odot}$ and 4 M$_{\odot}$, by increasing the abundance of
carbon. In stars more massive than 4 M$_{\odot}$, the HBB produces an
abrupt increase of nitrogen and a similar decrease of carbon. In
Figure~\ref{yields} we show the ejected masses of $^{12}$C and
$^{14}$N as a function of the stellar mass $M_{*}$, for $Z=$
Z$_{\odot}$, compared with those from RV, where we see that the
described behaviour for both elements does not appear since those
authors had not taken into account the transformation from carbon in
nitrogen during the HBB phase. In fact, nitrogen shows a variation
around 4~ M$_{\odot}$ for RV but it does not increase for larger
stellar masses.

\section{Computed models}

In order to compare the results obtained with these new stellar yields
with those produced by other sets, we have also used those from RV for
LIM stars. For massive stars, we have chosen those from Woosley \&
Weaver (1995; hereafter WW), who joined explosive nucleosynthesis to
the themselves pre-supernova yields for metallicities between $\rm Z =
0$ and $\rm Z = Z_{\odot}$; and those from Portinari, Chiosi \&
Bressan (1998; hereafter PCB), who obtained the yields for the Padova
group metallicities for $\rm M > 8$ M$_{\odot}$ and considered the
mass loss by stellar winds, the influence of the metallicity during
the stellar evolution and the explosive nucleosynthesis.  By combining
these sets, we run four models: BU-PCB, BU-WW, RV-PCB, RV-WW.

On the other hand, since the origin of nitrogen in massive stars is not
clear, we have also run the same four models but considering N as primary,
doing a total of eight models. They are used in the GCE model, as
implemented with the multiphase chemical evolution model (Ferrini et
al. 1992; Moll\'{a}, D\'{\i}az \& Ferrini 2002).

\section{Results}

Once the Galaxy is modeled, we study the behaviour of the CNO
elements. Results referring to gas structure, time evolution and
radial gradients are analyzed in Gavil\'{a}n, Buell \& Moll\'{a} (2003)

\begin{figure}[t]
\centerline{\psfig{file=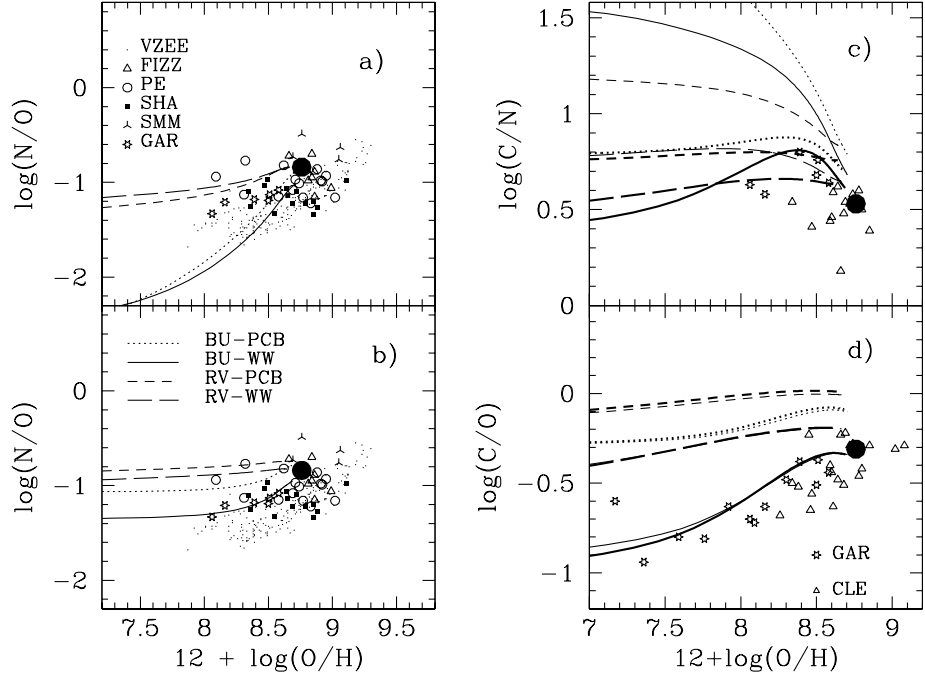,width=13cm,angle=-90}}
\caption{{\em Left:} The relation $\rm log(N/O)$, vs. $\rm 12
+log(O/H)$ whith a) Secondary nitrogen and b) Primary nitrogen. Data
are from Fitzsimmons et al. (1992, FIZZ), Garnett et al. (1999, GAR),
Peimbert (1979, PE), Shaver et al. (1983, SHA), Smartt et al.  (2001,
SM), and Van Zee (1998, VZEE). {\em Right:} C) The relation $\rm log(C/N)$
and d) $\rm log(C/O)$ vs. as $\rm 12 +log(O/H)$. Data are from
Garnett et al. (1995; 1999, GAR), and from Clegg et al. (1981,
CLE). Thick lines mean N primary while thin lines are models with N as
secondary. The large filled dot is the observed solar value.}
\label{no}
\end{figure}

We study the relation $ log(N/O)$ vs. $12 +log (O/H)$ in
Figure~\ref{no} for models with nitrogen as secondary (panel a) or
primary (panel b).  No model with N as secondary reproduces the
observed trend.  Models fit the data better if nitrogen is considered
primary, mostly the BU-WW model. In this case, the model line has a
slope close to zero at the beginning, and then develops a positive
slope, caused by the behaviour of nitrogen yields for stars of
different stellar masses: when stars with masses larger than 4
M$_{\odot}$ begin to evolve, the nitrogen production suffers a great
change. In the RV models this change cannot be seen because yields of N 
are quite uniform.

The relation C/N vs. O/H is shown in Figure~\ref{no} panel c), where the
8 models are included.  Again, data are better reproduced when nitrogen
is primary --thick lines-- than when nitrogen is secondary --thin
lines.  In particular, Model BU-WW is able to reproduce the variable
slope shown by the data, due to the evolution of LIM stars with masses
close to 4 M$_{\odot}$, after the primary nitrogen ejected by massive
stars. The same effect can not be obtained either with secondary
nitrogen or with primary nitrogen in other yields. The effect of HBB
and third dredge up processes is seen in this figure as the decreasing
of C/N at high oxygen abundance. This means that nitrogen appears while
C is not ejected yet by the less massive stars, 

Finally, in Figure~\ref{no}, panel d), we show the relation of carbon
abundance with the oxygen abundances, which does not change with the
different origin of the nitrogen. Again Model BU-WW fits the
HII region observations  better than the others. Model BU-PCB shows the same
trend but with a higher absolute value. This variable slope of C/O vs.
O/H is explained here by the normal evolution of LIM stars, instead of
using metallicity depending yields.

\section{Conclusions}

We have used new stellar yields based on a detailed modeling of the
TP-AGB phase, which affects the nitrogen and carbon production
directly, in a model of Chemical Evolution of the Galaxy disk. The
evolution of the solar neighborhood obtained with this model including
these yields for LIM stars and WW yields for massive stars, is able to
reproduce the HII regions and stellar data.  The trend of C/N vs. O/H
is reproduced without using metallicity depending yields.

\end{document}